\documentclass[sigconf,nonacm]{acmart}

\usepackage{graphicx}
\usepackage{booktabs}

\graphicspath{{figures/}}

\renewcommand\footnotetextcopyrightpermission[1]{}
\begin{document}

\title{FocusBuddy: Encouraging Healthy Desk-Work Habits by Caring for a Virtual Pet on a Water Bottle}

\author{Mohamed Ouf}
\affiliation{%
  \institution{Queen's University}
  \city{Kingston}
  \country{Canada}
}
\email{2blr2@queensu.ca}

\author{Rowan Hussein}
\affiliation{%
  \institution{University of Ottawa}
  \city{Ottawa}
  \country{Canada}
}
\email{rhuss060@uottawa.ca}

\renewcommand{\shortauthors}{Ouf and Hussein}

\begin{abstract}
People who study or work at a desk sit for long uninterrupted periods and drink less water than they intend to. Software reminders address both problems but are easy to dismiss and easy to resent. We present FocusBuddy, a proof-of-concept fabric case that wraps a standard water bottle and houses a microcontroller, environmental sensors, and a small display showing a virtual pet. The pet's condition mirrors the user's self-care: drinking water feeds the pet, standing up to move plays with it, and refilling an empty bottle cleans it. Twenty undergraduate students used FocusBuddy for two weeks during their regular coursework and completed a written interview. Self-reported water intake rose from a median of 3 to 4 cups per day, movement episodes rose from 2 to 4 per day, and the longest uninterrupted sitting period fell from 162.5 to 90 minutes (all $p \leq .001$). Written responses suggest that the pet framing helps mainly when prompts respect focused work, and that pet neglect can convert a wellness prompt into a source of guilt. We contribute the prototype, first-deployment evidence of healthy-direction shifts in self-reported habits, and design implications for emotionally framed wellness devices.
\end{abstract}

\keywords{virtual pet, gamification, tangible interaction, wellness, sedentary behavior, hydration, proof of concept}

\maketitle

\section{Introduction}
Prolonged sitting is associated with elevated metabolic risk independent of exercise~\cite{owen2010sitting, healy2008breaks}, and expert guidance for desk workers now recommends regularly breaking up seated time during the day~\cite{buckley2015sedentary}. Students and knowledge workers know this. The difficulty is not knowledge but timing and motivation: a reminder to stand up competes with the task at hand, lands at moments the software cannot judge~\cite{iqbal2008notification, mark2008interrupted}, and carries nothing beyond the reminder itself. Notification fatigue follows, and the reminders get dismissed unread.

One long-standing answer in HCI is to make the healthy action mean something beyond itself. Persuasive health systems have attached step counts to the growth of a garden~\cite{consolvo2008ubifit}, the wellbeing of a shared fish tank~\cite{lin2006fishnsteps}, and the feeding of a virtual pet~\cite{pollak2010timetoeat}. A separate line of work has moved health feedback off the phone and into everyday objects, from an ambient sculpture that slumps as its owner sits~\cite{jafarinaimi2005breakaway} to instrumented mugs and bottles that sense drinking~\cite{chiu2009playful, fortmann2014waterjewel}. These two ideas are rarely combined: the emotional pull of a creature to care for, embedded directly in the object the healthy behavior uses.

We combine them in FocusBuddy, a fabric case that wraps a standard 1-liter water bottle and houses a BBC micro:bit~\cite{austin2020microbit}, a battery base, two buttons, and an LED matrix showing a virtual pet. The pet's needs map onto the user's own: drinking, moving, and a comfortable room. Caring for the pet therefore requires caring for oneself. The bill of materials amounts to a hobby microcontroller and some fabric, which makes the concept testable without custom hardware.

This paper reports a proof-of-concept evaluation with 20 undergraduate students at two Canadian universities who used FocusBuddy during their regular coursework and completed a structured written interview. We contribute (1) the FocusBuddy prototype and its design rationale, (2) evidence that use coincided with self-reported shifts in hydration, movement, and uninterrupted sitting, and (3) qualitative findings on when a pet framing helps and when it backfires, with implications for emotionally framed wellness devices.

\section{Related Work}
Sedentary behavior carries health risks that accumulate through long unbroken sitting bouts rather than total sitting alone~\cite{owen2010sitting, healy2008breaks}, and workplace guidance recommends regular standing and movement breaks~\cite{buckley2015sedentary}. HCI systems that prompt such breaks confront the interruption problem: poorly timed prompts impose measurable cost and stress on the interrupted task~\cite{mark2008interrupted}, and deferring notifications to task boundaries reduces that cost~\cite{iqbal2008notification}. Break-prompting systems such as Time for Break~\cite{luo2018timeforbreak} and BreakSense~\cite{cambo2017breaksense} found that workers accept prompts selectively and negotiate them around their tasks. Our findings echo this: prompt timing was the strongest determinant of whether participants acted.

A second body of work motivates health behaviors through game elements~\cite{deterding2011gamification}. Reviews find gamified health interventions often work but with mixed and context-dependent effects~\cite{hamari2014does, johnson2016gamification, wattanasoontorn2013serious}. Within this space, caretaking metaphors recur: UbiFit Garden grew flowers from physical activity~\cite{consolvo2008ubifit}, Fish'n'Steps tied a fish's mood to step counts~\cite{lin2006fishnsteps}, and Time to Eat had participants feed a virtual pet by photographing healthy meals, which increased healthy eating among adolescents~\cite{pollak2010timetoeat}. Virtual companions can elicit attachment~\cite{chesney2007illusion}, and emotional engagement with objects shapes how people use them~\cite{norman2004emotional}. Gamification also has a documented dark side, including pressure and guilt when the mechanic punishes lapses~\cite{toda2018darkside, kim2016ethics}. Our study observed both faces of this mechanic in a single deployment.

A third thread embeds health feedback in tangible, peripheral objects rather than screens, following the ubiquitous-computing argument that calm technology should live at the periphery of attention~\cite{weiser1991computer, pousman2006taxonomy, ishii1997tangible}. Breakaway used a slumping desk sculpture to reflect sitting time~\cite{jafarinaimi2005breakaway}. For hydration specifically, Playful Bottle attached a phone to a mug and turned sensed drinking into the watering of a virtual tree~\cite{chiu2009playful}, and WaterJewel moved intake feedback onto a bracelet~\cite{fortmann2014waterjewel}. Playful Bottle comes closest to our design, but its creature lives on a phone docked to the mug and grows from drinking alone. FocusBuddy makes the container the permanent home of a Tamagotchi-style dependent pet~\cite{pollak2010timetoeat, chesney2007illusion} whose needs extend to movement and the surrounding room, so the object the user cares for is also the object they use to care for themselves.

\section{The FocusBuddy Prototype}

\subsection{Design Process}
\begin{figure}[tb]
  \centering
  \includegraphics[width=\linewidth]{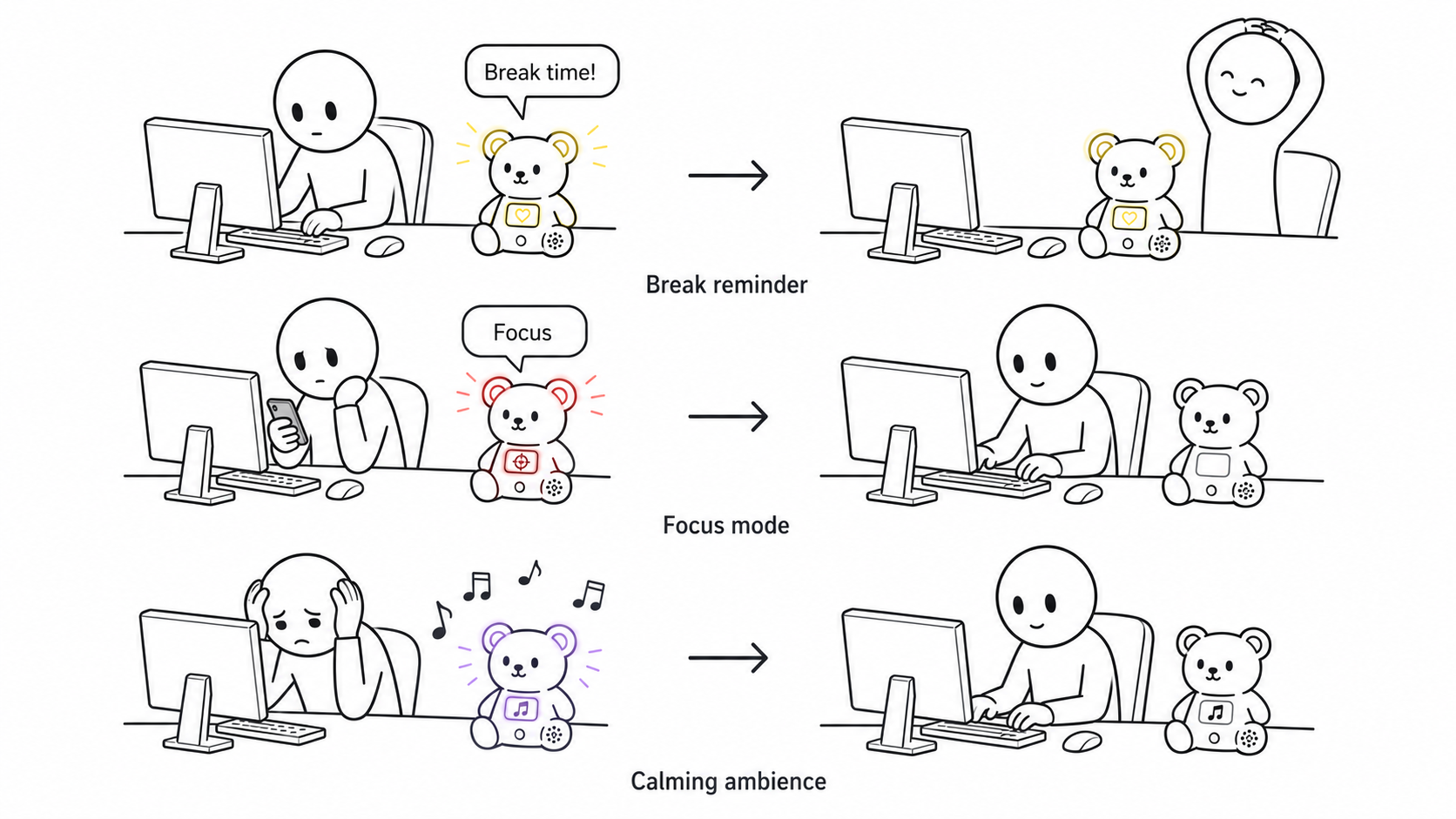}
  \Description{Three-panel storyboard. In each panel a bear-shaped desk device prompts a person at a computer: a break reminder, a focus prompt, and calming music, each followed by the person complying.}
  \caption{Storyboard of the original desk-companion concept: break prompts, focus prompts, and calming ambience from a bear-shaped device. Testers rejected the added desk object, which redirected the design onto the water bottle.}
  \label{fig:original-concept}
\end{figure}

\begin{figure}[tb]
  \centering
  \includegraphics[width=\linewidth]{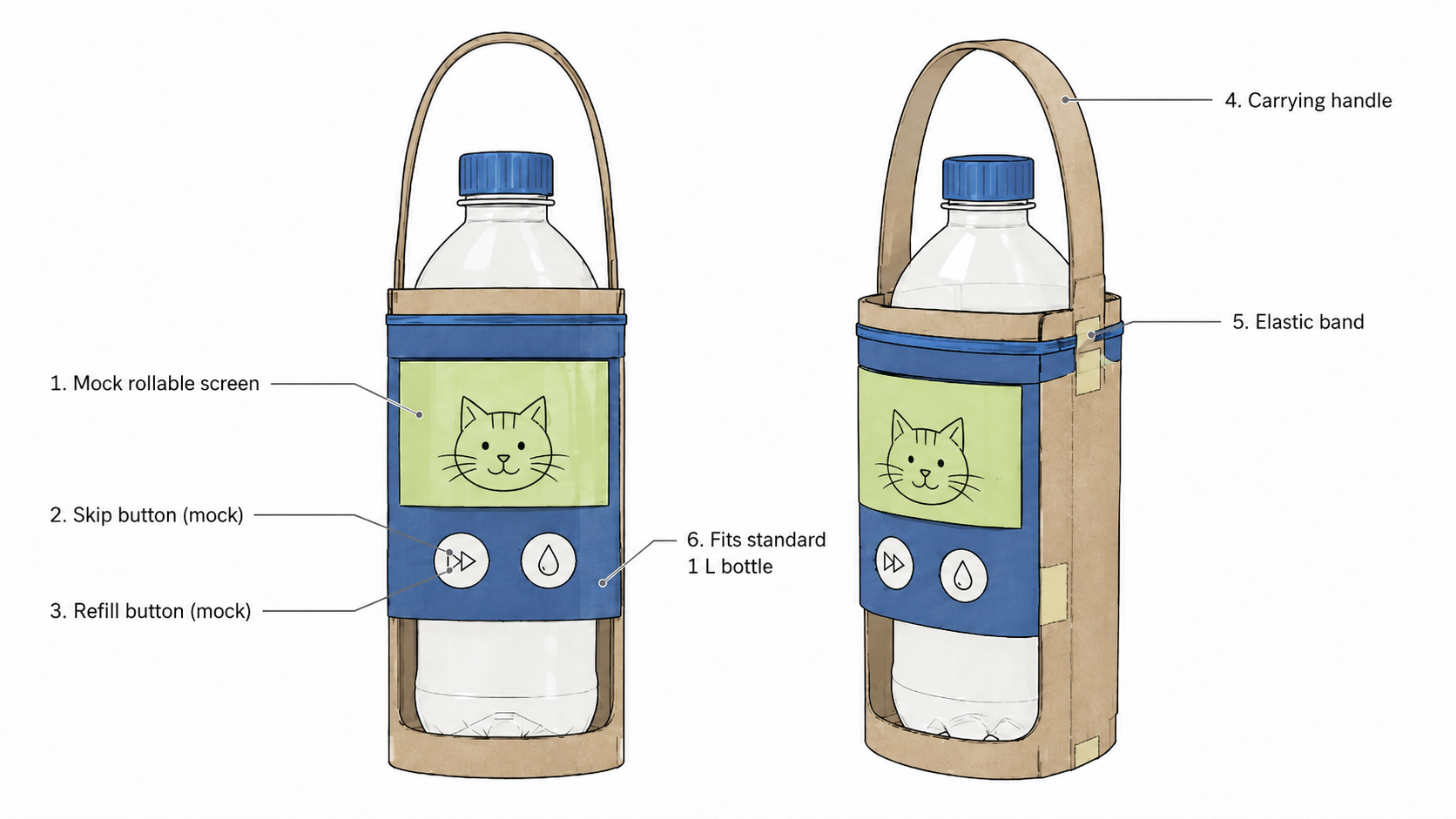}
  \Description{Two annotated views of a cardboard sleeve around a plastic water bottle, labeling the mock screen with a cat face, mock Skip and Refill buttons, carrying handle, and elastic band.}
  \caption{Mid-fidelity prototype: a paper-and-cardboard case with a mock screen and mock Skip and Refill buttons, used in Wizard-of-Oz sessions.}
  \label{fig:mid-fidelity}
\end{figure}

\begin{figure}[tb]
  \centering
  \includegraphics[width=0.8\linewidth]{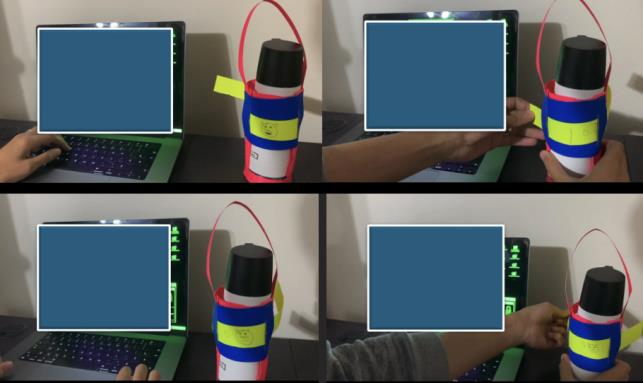}
  \Description{A tester at a desk drinks from a water bottle wearing the cardboard prototype case while a researcher operates its mock display.}
  \caption{A design-phase tester using the mid-fidelity prototype in a Wizard-of-Oz session, with a researcher simulating the display responses.}
  \label{fig:wizard-of-oz}
\end{figure}

Our first concept was a desk-bound companion toy, a small bear with a screen, LEDs, and speakers that would sit beside the monitor and issue wellness prompts (Figure~\ref{fig:original-concept}). Pilot feedback was negative on form factor rather than function: testers called a toy on a work desk unprofessional and resented carrying one more object. We kept the companion framing and moved it onto an object our target users already carry. We then tested a paper-and-cardboard mid-fidelity case (Figure~\ref{fig:mid-fidelity}) with a Wizard-of-Oz protocol~\cite{dahlback1993wizard} in which a researcher simulated the display (Figure~\ref{fig:wizard-of-oz}). These sessions confirmed that a bottle-mounted display with two buttons felt acceptable and legible, and that plain reminders without a game layer felt too easy to ignore. Following Houde and Hill~\cite{houde1997prototypes}, each prototype asked one question: the bear asked about role, the cardboard case about look and feel, and the final build about implementation.

\subsection{Final Prototype}
\begin{figure}[tb]
  \centering
  \includegraphics[width=0.85\linewidth]{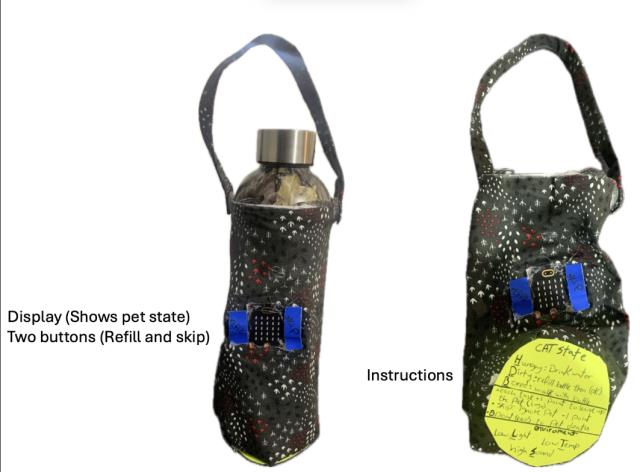}
  \Description{Photograph of the built prototype: a dark fabric sleeve around a water bottle with a small LED display, two buttons, and a yellow circular 3D-printed base.}
  \caption{The working FocusBuddy prototype: fabric sleeve, micro:bit display and sensors, Skip and Refill buttons, and 3D-printed battery base.}
  \label{fig:high-fidelity}
\end{figure}

The final prototype (Figure~\ref{fig:high-fidelity}) is a fabric sleeve that fits most 1-liter bottles, closed by an elastic band at the rim, with a removable handle and a 3D-printed base that houses the batteries. A BBC micro:bit~\cite{austin2020microbit} provides the 5$\times$5 LED matrix that displays the pet, plus an accelerometer, temperature sensor, light sensor, and microphone. Two labeled buttons, Skip and Refill, are the only controls.

\begin{figure}[tb]
  \centering
  \begin{minipage}{0.48\linewidth}
    \includegraphics[width=\linewidth]{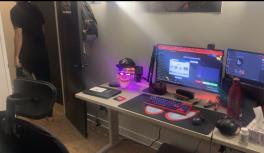}
  \end{minipage}
  \hfill
  \begin{minipage}{0.48\linewidth}
    \includegraphics[width=\linewidth]{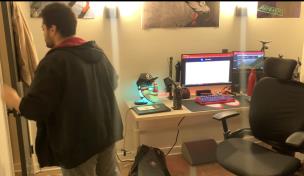}
  \end{minipage}
  \Description{Left: a person walks outdoors carrying the water bottle in its FocusBuddy case. Right: a person adjusts a desk lamp beside the bottle after an environment prompt.}
  \caption{The final prototype in use during pre-study testing sessions: taking the pet along on a walk (left) and adjusting the lighting after an environment prompt (right).}
  \label{fig:in-use}
\end{figure}

The pet, a small cat rendered on the LED matrix, has three needs on timers and one tied to the environment. Thirty minutes after the last detected drink the pet grows \textit{hungry}, and the accelerometer registers the bottle-tipping motion of drinking as feeding. An empty bottle makes the pet \textit{dirty}, cleared by refilling and pressing Refill. An hour without user movement makes the pet \textit{bored}, cleared when the accelerometer detects the bottle being carried on a walk or stretch (Figure~\ref{fig:in-use}, left). A dark, loud, or cold room makes the pet uncomfortable, prompting the user to adjust the room. Completed care actions add points and play a happy animation, ignored prompts subtract points, and at zero points the pet dies. Skip dismisses a prompt at a small point cost, a deliberate escape hatch for busy moments.

\section{Study}
We asked two questions. Does using FocusBuddy coincide with measurable change in self-reported hydration, movement, and sitting? And how do users experience a wellness prompt delivered as the need of a pet? A structured written interview suits a proof-of-concept at this stage: it captures both pre/post self-estimates and open-ended accounts from all participants at low burden~\cite{klasnja2011evaluate}.

\textbf{Participants.} Twenty undergraduate students took part, ten from Queen's University and ten from the University of Ottawa, aged 18 to 23 (median 21). All spent substantial weekday time at a desk: 4 reported 4--6 hours, 9 reported 6--8, 4 reported 8--10, and 3 reported over 10. The written interview did not collect gender or program of study. Each author recruited participants at their own university. Participation was voluntary, unpaid, and anonymous, and participants could skip any question or stop at any time.

\textbf{Procedure.} Each participant received a FocusBuddy unit and a one-page instruction sheet and used the device during their normal coursework for two weeks. Usage varied by choice: 6 participants used it on 10 or more days, 7 on 6--9 days, 5 on 3--5 days, and 2 on 0--2 days. We retained the low-use participants in all analyses since their reasons for lapsing are themselves findings. Afterward, participants completed a written interview with three parts. The first collected habit estimates for a typical workday before the study and during its last week: cups of water while working, times standing up to move or stretch, and longest uninterrupted sitting period in minutes. The second contained nine agreement items (1--5), one of them an embedded attention check, covering usability (UMUX-Lite~\cite{lewis2013umux}), enjoyment and pressure (adapted from the Intrinsic Motivation Inventory~\cite{mcauley1989imi}), and continued-use intention. The third posed five open-ended questions asking for specific incidents rather than opinions. All 20 participants passed the attention check, so no responses were excluded.

\textbf{Analysis.} We compared the three paired habit estimates with two-sided exact Wilcoxon signed-rank tests at a Bonferroni-corrected $\alpha = .0167$, reporting the matched-pairs rank-biserial correlation $r$ as effect size. The first author coded the open-ended answers inductively into themes. We report a theme when at least three participants expressed it. The interview instrument and anonymized responses are available at \url{https://github.com/MohamadAshrafSalama/focusbuddy-study-materials}.

\section{Findings}

\subsection{Self-Reported Behavior Change}
\begin{figure}[tb]
  \centering
  \includegraphics[width=\linewidth]{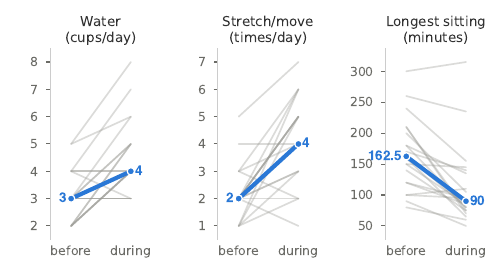}
  \Description{Three slope panels comparing each participant's before and during estimates for water intake, movement episodes, and longest sitting period. Most gray participant lines slope in the healthier direction, a few stay flat, and the blue median lines follow the same pattern.}
  \caption{Self-reported habits before the study and during its last week. Gray lines are individual participants ($n = 20$), blue lines and labels are medians.}
  \label{fig:results}
\end{figure}

All three habits shifted in the healthy direction (Figure~\ref{fig:results}). Median water intake while working rose from 3 to 4 cups per day ($W = 2$, $p = .001$, $r = .96$), median movement episodes rose from 2 to 4 per day ($W = 7$, $p < .001$, $r = .92$), and the median longest sitting period fell from 162.5 to 90 minutes ($W = 7$, $p < .001$, $r = .93$). In relative terms, mean water intake rose 45\%, mean movement episodes rose 86\%, and the mean longest sitting period fell 32\%. Effect sizes this close to the ceiling on coarse retrospective estimates signal that nearly every participant moved in the same direction, and we read them as direction rather than calibrated magnitude. Beyond the three habit measures, 16 of 20 participants changed something physical about their workspace in response to the environmental prompts, most often lighting.

The individual slopes in Figure~\ref{fig:results} carry a caution the medians hide. The two participants who used the device on 2 or fewer days reported no improvement on any measure, with flat water intake for both and, for P06, slightly less movement and longer sitting than before. One heavy-deadline participant (P14) reported drinking less and sitting longer than before. The shifts track engagement rather than mere possession of the device.

Agreement ratings (Table~\ref{tab:likert}) describe an experience that is easy but not yet compelling: ease of use was the strongest rating, requirements-fit and intention to keep using the device were moderate, and pressure from the pet's demands was low on average but reached the scale maximum for one participant. Written answers explain this spread.

\begin{table}[tb]
  \caption{The eight substantive agreement items (1 = strongly disagree, 5 = strongly agree), $n = 20$; the ninth item was the attention check. (R) marks reverse-framed items, reported unreversed.}
  \label{tab:likert}
  \begin{tabular}{@{}lc@{}}
    \toprule
    Item & Mean \\
    \midrule
    FocusBuddy is easy to use & 4.1 \\
    I paid more attention to my health habits & 3.6 \\
    I enjoyed taking care of the pet & 3.6 \\
    Capabilities meet my requirements & 3.5 \\
    I would keep using FocusBuddy & 3.4 \\
    Prompts interrupted my work at bad moments (R) & 2.9 \\
    I felt bad when the pet's condition got worse & 2.8 \\
    I felt pressure or stress from the pet's demands (R) & 2.5 \\
    \bottomrule
  \end{tabular}
\end{table}

\subsection{Timing Decides Whether Prompts Help}
The most common complaint, raised by 8 of 20 participants, was prompts that landed mid-focus: during a presentation (P06), minutes into an online meeting (P02), or while solving a hard problem (P12, P18). Participants did not reject the prompts themselves. P18 ignored two prompts while working through a difficult problem, then acted on a third that arrived just after the fix: ``that timing made it easy to stand up and refill my bottle; I probably would have kept sitting otherwise.'' What they rejected was the cost of the available responses. Skip resolved a prompt instantly but visibly saddened the pet, which, as P01 put it, ``made a ten-minute delay feel like a penalty rather than a useful option.'' The most common improvement request was some form of respecting focus: a snooze that leaves the pet unchanged (P01), a reschedule with visible confirmation (P12), calendar or quiet-hours integration (P02, P06), a deadline mode with batched prompts (P14), and a grace period after a completed break (P02, P18).

\subsection{The Pet Cuts Both Ways}
For engaged participants the pet reframed self-care as a relationship. P20 described the loop as reciprocal, ``I was helping it by helping myself,'' and P03 found the reminders ``caring rather than corrective, which mattered when I was already hard on myself.'' P05 described a midnight water prompt that ended three unbroken hours at the desk, and P13 found that a followed reminder ``also gave me emotional space before starting the next task.''

The same mechanism produced guilt when life got busy. P16, who rated pet-related pressure 5 of 5, took breaks ``mostly because I felt bad rather than because I wanted one,'' and noted that one missed prompt damaged the pet more visibly than a completed action repaired it. P13 read the sad pet as ``a reprimand'' on stressful days and avoided looking at the device. A third group felt nothing: P11 called the pet theme unappealing and wanted ``the straightforward functionality,'' and P02 preferred a neutral configurable timer. Three participants (P13, P16, P02) independently proposed the same fix: let users decouple the wellness prompts from the pet's mood, or select a neutral feedback style.

\subsection{Trust in Sensing}
A quieter theme concerned confirmation. Participants who acted on a prompt sometimes could not tell whether the device had noticed: a refill that produced no reaction (P20), a walk to refill the bottle that the device treated as a missed break (P18), and ambiguous status icons (P03). P08 wanted a private weekly summary of accepted and skipped prompts to judge whether the habit was actually forming. Unacknowledged care actions erode exactly the relationship the pet is meant to build.

\subsection{Legibility of a Tiny Pet}
Four participants struggled to read the 5$\times$5 pet at the moment a prompt mattered. P03 found the status icons ambiguous: ``I could not tell whether I had missed a prompt, the bottle action had not registered, or I needed to do something else.'' Animation cost time that glancing users did not have. P05 wanted a dim late-night check-in but ``the animation took long enough that I sometimes closed it before seeing what it asked,'' P15 found the small labels hard to read in a dim room, and P11 reported that the celebratory reward visual took over the display and hid the next reminder time. The pet's expressiveness, already limited by the LED matrix, competed with the display's second job of stating plainly what the device wants.

\subsection{Fitting Shared Spaces and Bags}
FocusBuddy travels with the bottle, and three participants hit the limits of that portability. P07 skipped suggested stretches in shared study spaces ``rather than draw attention to myself,'' P18 found the sound cue ``awkward in the library,'' and P17 left the case behind in the library: ``it felt bulky to carry, and I worried that the attachment would catch on my bag or bottle.'' A device that lives on an everyday object inherits every social setting that object enters, and the current prototype only fits some of them.

\section{Discussion}

\subsection{Proof of Concept, and What It Justifies}
FocusBuddy set out to test whether a virtual pet living on a water bottle can move desk-work habits, using nothing more than a hobby microcontroller and fabric. The self-reports say yes: all three targeted habits shifted with large effect sizes, and 16 of 20 participants also adjusted their physical workspace. We claim only that the design shows promise. The measures are retrospective self-estimates, there was no control condition, and a short deployment cannot separate the pet's contribution from novelty or from being reminded at all~\cite{klasnja2011evaluate}. The gradient in our data, where benefit tracked use, mirrors a pattern familiar from other voluntary-participation settings: in open-source communities, early participation experiences relate to whether newcomers persist and advance to central roles~\cite{ouf2026dogood, ouf2026sameproject}. What the study supports is the next, better-instrumented iteration, and the written answers sharpen what that iteration must fix.

\subsection{Emotional Stakes Need an Asymmetry}
The pet framing worked through attachment, as virtual-companion research predicts~\cite{chesney2007illusion, pollak2010timetoeat}, but attachment set the stakes for failure too. Participants under deadline pressure experienced the declining pet as an accusation, matching the pressure and guilt effects documented for punishing gamification mechanics~\cite{toda2018darkside, kim2016ethics}. The asymmetry P16 observed, where one lapse visibly outweighed one good action, is a design error rather than an inherent cost of the metaphor. A companion that forgives lapses quickly but celebrates care conspicuously would keep the warm loop our engaged participants described and remove the reprimand our stressed participants avoided. Making the emotional layer optional, as three participants requested, acknowledges that attachment is a preference rather than a universal motivator. Designs that treat users as one uniform population blur these differences, a caution prior work has raised for other communities that split into distinct sub-genres with distinct needs~\cite{ouf2026typology, ouf2026oss4sg}.

\subsection{Prompts Must Defer to Focus}
Our participants' near-unanimous request (snooze without penalty, quiet hours, post-break grace periods) restates the interruptibility literature~\cite{iqbal2008notification, mark2008interrupted, luo2018timeforbreak} in the vocabulary of pet care: the pet should trust its owner to come back. The accelerometer, microphone, and light sensor already on board could approximate focus and meeting detection cheaply, deferring non-urgent needs to activity boundaries.

\subsection{Limitations}
Beyond the self-report and control limitations above, our sample is 20 undergraduates at two Canadian universities, so the results may not transfer to office workers, and the before-study estimates were collected in the same end-of-study interview as the during-study estimates, inviting recall and demand biases. The written-interview format traded probing depth for coverage. Coding was performed by a single author. The prototype itself constrained the experience: a 5$\times$5 LED matrix limits the pet's expressiveness, and several participants found the icons ambiguous.

\subsection{Future Work}
The concept now deserves hardware that removes the prototype's two bottlenecks: expressiveness and sensing confirmation. Compact microcontroller boards that pair a thin high-resolution display with an inertial unit and environmental sensors would allow a legible, animated pet, explicit acknowledgment of each care action, and on-device activity inference, in a package thinner than the current micro:bit stack. Sensing-and-inference stacks of this kind are maturing across application domains~\cite{soliman2024digitaltwin, marshall2024ris}. Richer sensing could also borrow from adjacent work: vision-based systems assess movement form during exercise~\cite{ouf2024visionpf}, single-channel EEG estimates cognitive load in learning settings~\cite{hussein2026eeg}, and language models recover user-facing intent from raw artifacts~\cite{ouf2025userstories}. These are candidate routes to focus detection, verified movement episodes, and prompts worded to the user's current task. With that platform we plan a longer counterbalanced deployment with logged (rather than recalled) drinking and movement, a neutral-reminder control condition, and office workers alongside students, testing directly whether the pet earns its emotional stakes.

\section{Conclusion}
FocusBuddy embeds a needy virtual pet in a water bottle case so that caring for the pet and caring for oneself are the same actions. Twenty undergraduates who lived with the prototype reported drinking more, moving more, and sitting in shorter stretches, and their written accounts locate both the power and the risk of the design in the same place: the pet's feelings. When prompts respected focus, the pet turned self-care into a small reciprocal relationship. When they did not, it manufactured guilt. A proof of concept this cheap that moves behavior this much argues that the tangible pet deserves a serious second iteration, one that forgives its owner as readily as it asks to be fed.

\bibliographystyle{ACM-Reference-Format}
\bibliography{references}

@article{owen2010sitting,
  author  = {Owen, Neville and Healy, Genevieve N. and Matthews, Charles E. and Dunstan, David W.},
  title   = {Too Much Sitting: The Population Health Science of Sedentary Behavior},
  journal = {Exercise and Sport Sciences Reviews},
  year    = {2010},
  volume  = {38},
  number  = {3},
  pages   = {105--113},
  doi     = {10.1097/JES.0b013e3181e373a2}
}

@article{healy2008breaks,
  author  = {Healy, Genevieve N. and Dunstan, David W. and Salmon, Jo and Cerin, Ester and Shaw, Jonathan E. and Zimmet, Paul Z. and Owen, Neville},
  title   = {Breaks in Sedentary Time: Beneficial Associations with Metabolic Risk},
  journal = {Diabetes Care},
  year    = {2008},
  volume  = {31},
  number  = {4},
  pages   = {661--666},
  doi     = {10.2337/dc07-2046}
}

@article{buckley2015sedentary,
  author  = {Buckley, John P. and Hedge, Alan and Yates, Thomas and Copeland, Robert J. and Loosemore, Michael and Hamer, Mark and Bradley, Gavin and Dunstan, David W.},
  title   = {The Sedentary Office: An Expert Statement on the Growing Case for Change Towards Better Health and Productivity},
  journal = {British Journal of Sports Medicine},
  year    = {2015},
  volume  = {49},
  number  = {21},
  pages   = {1357--1362},
  doi     = {10.1136/bjsports-2015-094618}
}

@inproceedings{iqbal2008notification,
  author    = {Iqbal, Shamsi T. and Bailey, Brian P.},
  title     = {Effects of Intelligent Notification Management on Users and Their Tasks},
  booktitle = {Proceedings of the SIGCHI Conference on Human Factors in Computing Systems (CHI '08)},
  year      = {2008},
  pages     = {93--102},
  publisher = {ACM},
  doi       = {10.1145/1357054.1357070}
}

@inproceedings{mark2008interrupted,
  author    = {Mark, Gloria and Gudith, Daniela and Klocke, Ulrich},
  title     = {The Cost of Interrupted Work: More Speed and Stress},
  booktitle = {Proceedings of the SIGCHI Conference on Human Factors in Computing Systems (CHI '08)},
  year      = {2008},
  pages     = {107--110},
  publisher = {ACM},
  doi       = {10.1145/1357054.1357072}
}

@inproceedings{luo2018timeforbreak,
  author    = {Luo, Yuhan and Lee, Bongshin and Wohn, Donghee Yvette and Rebar, Amanda L. and Conroy, David E. and Choe, Eun Kyoung},
  title     = {Time for Break: Understanding Information Workers' Sedentary Behavior Through a Break Prompting System},
  booktitle = {Proceedings of the 2018 CHI Conference on Human Factors in Computing Systems (CHI '18)},
  year      = {2018},
  pages     = {1--12},
  publisher = {ACM},
  doi       = {10.1145/3173574.3173701}
}

@inproceedings{cambo2017breaksense,
  author    = {Cambo, Scott A. and Avrahami, Daniel and Lee, Matthew L.},
  title     = {BreakSense: Combining Physiological and Location Sensing to Promote Mobility During Work-Breaks},
  booktitle = {Proceedings of the 2017 CHI Conference on Human Factors in Computing Systems (CHI '17)},
  year      = {2017},
  pages     = {3595--3607},
  publisher = {ACM},
  doi       = {10.1145/3025453.3026021}
}

@inproceedings{deterding2011gamification,
  author    = {Deterding, Sebastian and Dixon, Dan and Khaled, Rilla and Nacke, Lennart E.},
  title     = {From Game Design Elements to Gamefulness: Defining ``Gamification''},
  booktitle = {Proceedings of the 15th International Academic MindTrek Conference: Envisioning Future Media Environments (MindTrek '11)},
  year      = {2011},
  pages     = {9--15},
  publisher = {ACM},
  doi       = {10.1145/2181037.2181040}
}

@inproceedings{hamari2014does,
  author    = {Hamari, Juho and Koivisto, Jonna and Sarsa, Harri},
  title     = {Does Gamification Work? {A} Literature Review of Empirical Studies on Gamification},
  booktitle = {Proceedings of the 47th Hawaii International Conference on System Sciences (HICSS)},
  year      = {2014},
  pages     = {3025--3034},
  publisher = {IEEE},
  doi       = {10.1109/HICSS.2014.377}
}

@article{johnson2016gamification,
  author  = {Johnson, Daniel and Deterding, Sebastian and Kuhn, Kerri-Ann and Staneva, Aleksandra and Stoyanov, Stoyan and Hides, Leanne},
  title   = {Gamification for Health and Wellbeing: A Systematic Review of the Literature},
  journal = {Internet Interventions},
  year    = {2016},
  volume  = {6},
  pages   = {89--106},
  doi     = {10.1016/j.invent.2016.10.002}
}

@article{wattanasoontorn2013serious,
  author  = {Wattanasoontorn, Voravika and Boada, Imma and Garc{\'i}a, Rub{\'e}n and Sbert, Mateu},
  title   = {Serious Games for Health},
  journal = {Entertainment Computing},
  year    = {2013},
  volume  = {4},
  number  = {4},
  pages   = {231--247},
  doi     = {10.1016/j.entcom.2013.09.002}
}

@article{kim2016ethics,
  author  = {Kim, Tae Wan and Werbach, Kevin},
  title   = {More Than Just a Game: Ethical Issues in Gamification},
  journal = {Ethics and Information Technology},
  year    = {2016},
  volume  = {18},
  number  = {2},
  pages   = {157--173},
  doi     = {10.1007/s10676-016-9401-5}
}

@incollection{toda2018darkside,
  author    = {Toda, Armando Maciel and Valle, Pedro Henrique Dias and Isotani, Seiji},
  title     = {The Dark Side of Gamification: An Overview of Negative Effects of Gamification in Education},
  booktitle = {Higher Education for All. From Challenges to Novel Technology-Enhanced Solutions (HEFA 2017)},
  series    = {Communications in Computer and Information Science},
  year      = {2018},
  pages     = {143--156},
  publisher = {Springer},
  doi       = {10.1007/978-3-319-97934-2_9}
}

@inproceedings{consolvo2008ubifit,
  author    = {Consolvo, Sunny and McDonald, David W. and Toscos, Tammy and Chen, Mike Y. and Froehlich, Jon and Harrison, Beverly and Klasnja, Predrag and LaMarca, Anthony and LeGrand, Louis and Libby, Ryan and Smith, Ian and Landay, James A.},
  title     = {Activity Sensing in the Wild: A Field Trial of {UbiFit Garden}},
  booktitle = {Proceedings of the SIGCHI Conference on Human Factors in Computing Systems (CHI '08)},
  year      = {2008},
  pages     = {1797--1806},
  publisher = {ACM},
  doi       = {10.1145/1357054.1357335}
}

@inproceedings{lin2006fishnsteps,
  author    = {Lin, James J. and Mamykina, Lena and Lindtner, Silvia and Delajoux, Gregory and Strub, Henry B.},
  title     = {Fish'n'Steps: Encouraging Physical Activity with an Interactive Computer Game},
  booktitle = {UbiComp 2006: Ubiquitous Computing, 8th International Conference},
  series    = {Lecture Notes in Computer Science},
  volume    = {4206},
  year      = {2006},
  pages     = {261--278},
  publisher = {Springer},
  doi       = {10.1007/11853565_16}
}

@inproceedings{jafarinaimi2005breakaway,
  author    = {Jafarinaimi, Nassim and Forlizzi, Jodi and Hurst, Amy and Zimmerman, John},
  title     = {Breakaway: An Ambient Display Designed to Change Human Behavior},
  booktitle = {CHI '05 Extended Abstracts on Human Factors in Computing Systems},
  year      = {2005},
  pages     = {1945--1948},
  publisher = {ACM},
  doi       = {10.1145/1056808.1057063}
}

@article{pollak2010timetoeat,
  author  = {Pollak, John P. and Gay, Geri and Byrne, Sahara and Wagner, Emily and Retelny, Daniela and Humphreys, Lee},
  title   = {It's Time to Eat! {U}sing Mobile Games to Promote Healthy Eating},
  journal = {IEEE Pervasive Computing},
  year    = {2010},
  volume  = {9},
  number  = {3},
  pages   = {21--27},
  doi     = {10.1109/MPRV.2010.41}
}

@inproceedings{chiu2009playful,
  author    = {Chiu, Meng-Chieh and Chang, Shih-Ping and Chang, Yu-Chen and Chu, Hao-Hua and Chen, Cheryl Chia-Hui and Hsiao, Fei-Hsiu and Ko, Ju-Chun},
  title     = {Playful Bottle: A Mobile Social Persuasion System to Motivate Healthy Water Intake},
  booktitle = {Proceedings of the 11th International Conference on Ubiquitous Computing (UbiComp '09)},
  year      = {2009},
  pages     = {185--194},
  publisher = {ACM},
  doi       = {10.1145/1620545.1620574}
}

@inproceedings{fortmann2014waterjewel,
  author    = {Fortmann, Jutta and Cobus, Vanessa and Heuten, Wilko and Boll, Susanne},
  title     = {WaterJewel: Design and Evaluation of a Bracelet to Promote a Better Drinking Behaviour},
  booktitle = {Proceedings of the 13th International Conference on Mobile and Ubiquitous Multimedia (MUM '14)},
  year      = {2014},
  pages     = {58--67},
  publisher = {ACM},
  doi       = {10.1145/2677972.2677976}
}

@article{chesney2007illusion,
  author  = {Chesney, Thomas and Lawson, Shaun},
  title   = {The Illusion of Love: Does a Virtual Pet Provide the Same Companionship as a Real One?},
  journal = {Interaction Studies},
  year    = {2007},
  volume  = {8},
  number  = {2},
  pages   = {337--342},
  doi     = {10.1075/is.8.2.09che}
}

@book{norman2004emotional,
  author    = {Norman, Donald A.},
  title     = {Emotional Design: Why We Love (or Hate) Everyday Things},
  publisher = {Basic Books},
  address   = {New York},
  year      = {2004}
}

@article{weiser1991computer,
  author  = {Weiser, Mark},
  title   = {The Computer for the 21st Century},
  journal = {Scientific American},
  year    = {1991},
  volume  = {265},
  number  = {3},
  pages   = {94--104},
  doi     = {10.1038/scientificamerican0991-94}
}

@inproceedings{pousman2006taxonomy,
  author    = {Pousman, Zachary and Stasko, John},
  title     = {A Taxonomy of Ambient Information Systems: Four Patterns of Design},
  booktitle = {Proceedings of the Working Conference on Advanced Visual Interfaces (AVI '06)},
  year      = {2006},
  pages     = {67--74},
  publisher = {ACM},
  doi       = {10.1145/1133265.1133277}
}

@inproceedings{ishii1997tangible,
  author    = {Ishii, Hiroshi and Ullmer, Brygg},
  title     = {Tangible Bits: Towards Seamless Interfaces Between People, Bits and Atoms},
  booktitle = {Proceedings of the ACM SIGCHI Conference on Human Factors in Computing Systems (CHI '97)},
  year      = {1997},
  pages     = {234--241},
  publisher = {ACM},
  doi       = {10.1145/258549.258715}
}

@article{austin2020microbit,
  author  = {Austin, Jonny and Baker, Howard and Ball, Thomas and Devine, James and Finney, Joe and de Halleux, Peli and H{\o}dneb{\o}, Steve and Moskal, Micha{\l} and Protzenko, Jonathan and Riley, Matt},
  title   = {The {BBC} micro:bit: From the {U.K.} to the World},
  journal = {Communications of the ACM},
  year    = {2020},
  volume  = {63},
  number  = {3},
  pages   = {62--69},
  doi     = {10.1145/3368856}
}

@article{dahlback1993wizard,
  author  = {Dahlb{\"a}ck, Nils and J{\"o}nsson, Arne and Ahrenberg, Lars},
  title   = {Wizard of {Oz} Studies --- Why and How},
  journal = {Knowledge-Based Systems},
  year    = {1993},
  volume  = {6},
  number  = {4},
  pages   = {258--266},
  doi     = {10.1016/0950-7051(93)90017-N}
}

@incollection{houde1997prototypes,
  author    = {Houde, Stephanie and Hill, Charles},
  title     = {What Do Prototypes Prototype?},
  booktitle = {Handbook of Human-Computer Interaction},
  edition   = {2nd},
  editor    = {Helander, Martin G. and Landauer, Thomas K. and Prabhu, Prasad V.},
  publisher = {Elsevier},
  year      = {1997},
  pages     = {367--381},
  doi       = {10.1016/B978-044481862-1.50082-0}
}

@inproceedings{klasnja2011evaluate,
  author    = {Klasnja, Predrag and Consolvo, Sunny and Pratt, Wanda},
  title     = {How to Evaluate Technologies for Health Behavior Change in {HCI} Research},
  booktitle = {Proceedings of the SIGCHI Conference on Human Factors in Computing Systems (CHI '11)},
  year      = {2011},
  pages     = {3063--3072},
  publisher = {ACM},
  doi       = {10.1145/1978942.1979396}
}

@inproceedings{lewis2013umux,
  author    = {Lewis, James R. and Utesch, Brian S. and Maher, Deborah E.},
  title     = {{UMUX-LITE}: When There's No Time for the {SUS}},
  booktitle = {Proceedings of the SIGCHI Conference on Human Factors in Computing Systems (CHI '13)},
  year      = {2013},
  pages     = {2099--2102},
  publisher = {ACM},
  doi       = {10.1145/2470654.2481287}
}

@article{mcauley1989imi,
  author  = {McAuley, Edward and Duncan, Terry E. and Tammen, Vance V.},
  title   = {Psychometric Properties of the Intrinsic Motivation Inventory in a Competitive Sport Setting: A Confirmatory Factor Analysis},
  journal = {Research Quarterly for Exercise and Sport},
  year    = {1989},
  volume  = {60},
  number  = {1},
  pages   = {48--58},
  doi     = {10.1080/02701367.1989.10607413}
}

@inproceedings{ouf2024visionpf,
  author    = {Ouf, Mohamed and Amin, R. M. and Sabry, M. M. and Ali, A. H. and Hamdan, Y. A. and Saleh, Sherine Nagy},
  title     = {{VISION-PF}: A Computer Vision Approach to Injury Prevention in Physical Fitness},
  booktitle = {2024 International Telecommunications Conference (ITC-Egypt)},
  year      = {2024},
  pages     = {1--6},
  publisher = {IEEE}
}

@inproceedings{ouf2025userstories,
  author    = {Ouf, Mohamed and Li, Haoyu and Zhang, M. and Guizani, Mariam},
  title     = {Reverse Engineering User Stories from Code Using Large Language Models},
  booktitle = {2025 IEEE International Conference on Collaborative Advances in Software and COmputing (CASCON)},
  year      = {2025},
  publisher = {IEEE}
}

@misc{ouf2026dogood,
  author        = {Ouf, Mohamed and Mohamed, Amr and Guizani, Mariam},
  title         = {Do Good, Stay Longer? {T}emporal Patterns and Predictors of Newcomer-to-Core Transitions in Conventional {OSS} and {OSS4SG}},
  year          = {2026},
  eprint        = {2601.23142},
  archivePrefix = {arXiv}
}

@misc{ouf2026oss4sg,
  author        = {Ouf, Mohamed and Noei, Shayan and Van Iterson, Zeph and Guizani, Mariam and Zou, Ying},
  title         = {An Empirical Analysis of Community and Coding Patterns in {OSS4SG} vs. Conventional {OSS}},
  year          = {2026},
  eprint        = {2601.03430},
  archivePrefix = {arXiv}
}

@article{soliman2024digitaltwin,
  author  = {Soliman, Abdelmoneim and Marshall, Mervin A. and Rahaman, Md Safiqur and Ouf, Mohamed and El-Sayed, Ahmed},
  title   = {A Bibliometric Review of Research Publications on Digital Twin Predictive Maintenance Systems in the Maritime Industry},
  journal = {Journal of Ocean Technology},
  year    = {2024},
  volume  = {19},
  number  = {2}
}

@article{marshall2024ris,
  author  = {Marshall, Mervin A. and Ouf, Mohamed and Elkhateeb, Abdelrahman and Soliman, Abdelmoneim and Rahaman, Md Safiqur and Mobarak, Sara},
  title   = {A Review of Reconfigurable Intelligent Surfaces and Their Application to Machine Learning-Assisted Underwater Communications},
  journal = {Journal of Ocean Technology},
  year    = {2024},
  volume  = {19},
  number  = {4}
}

@misc{ouf2026sameproject,
  author        = {Ouf, Mohamed and Guizani, Mariam},
  title         = {Same Project, Different Start: How Contribution Events Shape Activity and Retention in Open Source},
  year          = {2026},
  eprint        = {2604.22120},
  archivePrefix = {arXiv}
}

@misc{ouf2026typology,
  author        = {Ouf, Mohamed and Hussein, Rowan},
  title         = {Open Source Is Not One Thing: A Typology of Open-Source Software Sub-Genres},
  year          = {2026},
  eprint        = {2607.01750},
  archivePrefix = {arXiv}
}

@misc{hussein2026eeg,
  author        = {Hussein, Rowan and Ouf, Mohamed},
  title         = {Single-Channel {EEG}-Based Cognitive Load Assessment in Online Learning: A Hybrid Deep Learning Approach},
  year          = {2026},
  eprint        = {2607.01795},
  archivePrefix = {arXiv}
}

\end{document}